\documentclass[10pt, conference]{IEEEtran}

\usepackage{url}
\usepackage{cite}
\usepackage{amssymb}
\usepackage{bbding}
\usepackage{color, colortbl}
\definecolor{Gray}{gray}{0.95}
\definecolor{caribbeangreen}{rgb}{0.0, 0.8, 0.6}
\definecolor{carminepink}{rgb}{0.92, 0.3, 0.26}
\usepackage{booktabs}
\usepackage{array}
\usepackage[utf8]{inputenc}
\usepackage[T1]{fontenc}
\usepackage{microtype}
\usepackage{tabularx}
\usepackage{makecell}

\usepackage{amsmath}
\usepackage{lipsum}

\ifCLASSINFOpdf
   \usepackage[pdftex]{graphicx}

\else

\fi

\usepackage{todonotes}

\begin{document}
\title{On the Emotion of Users in App Reviews}


\author{\IEEEauthorblockN{Daniel Martens}
\IEEEauthorblockA{University of Hamburg\\
Hamburg, Germany\\
martens@informatik.uni-hamburg.de}
\and
\IEEEauthorblockN{Timo Johann}
\IEEEauthorblockA{University of Hamburg\\
Hamburg, Germany\\
johann@informatik.uni-hamburg.de}
}


\maketitle


\begin{abstract}
App store analysis has become an important discipline in recent software engineering research. It empirically studies apps using information mined from their distribution platforms. Information provided by users, such as app reviews, are of high interest to developers. Commercial providers such as App Annie analyzing this information became an important source for companies developing and marketing mobile apps. In this paper, we perform an exploratory study, which analyzes over seven million reviews from the Apple AppStore regarding their emotional sentiment. Since recent research in this field used sentiments to detail and refine their results, we aim to gain deeper insights into the nature of sentiments in user reviews. In this study we try to evaluate whether or not the emotional sentiment can be an informative feature for software engineers, as well as pitfalls of its usage. We present our initial results and discuss how they can be interpreted from the software engineering perspective.
\end{abstract}


\IEEEpeerreviewmaketitle


\section{Introduction}
\label{sec:intro}

In the past decade, app stores like Google Play or Apple AppStore became an important source for software engineers. Initially intended as software distribution platforms, their social aspects, such as the possibility for users to rate and review apps gained attention in the community. User reviews include useful information such as bug reports, feature requests or user experience \cite{pagano:RE:2013}. These reviews have been evaluated in different ways, including general exploratory studies \cite{pagano:RE:2013}, classification \cite{maalej:RE:2015}, feature extraction \cite{Guzman:RE:2014}, review filtering \cite{Chen:2014:AMI:2568225.2568263}, and summarization \cite{Fu:ICKDDM:2013}. However, user reviews are not mundane technical descriptions but also contain emotion. Many of the studies additionally applied sentiment analysis tools or APIs to improve their results. It remains unclear if existing tools are suitable for software engineering purposes \cite{Jongeling:2015:ICSME, Jongeling:2017:ESE}. For this reason, we present this exploratory study, and focus on the sentiment, to find out if positive or negative emotions in user reviews can be used as an additional feature in app store analysis. We intend to discover links between informational value and the emotion of user reviews. A special focus is placed on monitoring the evolution of the sentiment over time, to discover patterns and to draw conclusions by manually analyzing these patterns.

The remainder of this paper is organized as follows. Section \ref{sec:rel-work} presents related work in which emotion and sentiment has been used within the field of software engineering. In Section \ref{sec:design} we describe the design of the study, the collection and preparation of data, as well as the data analysis techniques used. In Section \ref{sec:results} we provide a brief descriptive overview of the data set, present initial findings, and draw conclusions from the software engineering perspective. We conclude with a discussion and directions for future work in Section \ref{sec:conclusion}.


\section{Related Work}
\label{sec:rel-work}

Natural language processing has been used for sentiment analysis for more than a decade \cite{Turney:2002:Corr}. With the rise of social media it gained increasing importance. Initially intended for the use in marketing or political opinion mining, the topic became popular in many other social domains. This also applied for social aspects of software engineering as the community recognized its potential. Most of the research has been conducted in the field of developers' interactions within bug trackers, commit messages or software related Q\&A sites \cite{Novielli:2014:SSE, Guzman:2014:SAC}. With the growing interest in app store mining \cite{Chen:2014:AMI:2568225.2568263, Harman:MSR:2012, Iacob:MSR:2013}, emotional sentiment has been used in research related to user reviews in app stores. In this area, sentiment analysis mostly played a supportive role to detail findings or to increase the quality of results. Goul et al.  used sentiment analysis on user reviews to address current bottlenecks in requirements\cite{Goul:2012:HICSS}. Li et al. calculated the overall user satisfaction in the app stores \cite{li:2010:user}. A more fine-grained study by Guzman and Maalej used sentiment analysis to summarize the opinion on app features retrieved from user reviews \cite{Guzman:RE:2014}. Maalej et al. added sentiments as an informative feature to train a classifier used to categorize user reviews regarding different topics \cite{maalej:RE:2015, maalej:2016:JRE}. However, Jongeling et al. reported on possible negative results when using sentiment analysis tools for software engineering research \cite{Jongeling:2017:ESE}. The use of sentiment analysis tools and APIs is highly dependent on the use case, and those have a significant impact on study results \cite{Jongeling:2015:ICSME}. Fu et al. \cite{Fu:ICKDDM:2013} built a regression model of words based on common user vocabulary and identified words with outstanding positive and negative sentiment based on their sample. Our study differentiates from the earlier studies therein that we undertake an isolated look on sentiments of user reviews in order to better understand if and how the sentiment can be used as a supportive feature for app store analyses.


\section{Study Design}
\label{sec:design}

\begin{figure}[t]
	\includegraphics[width=0.5\textwidth]{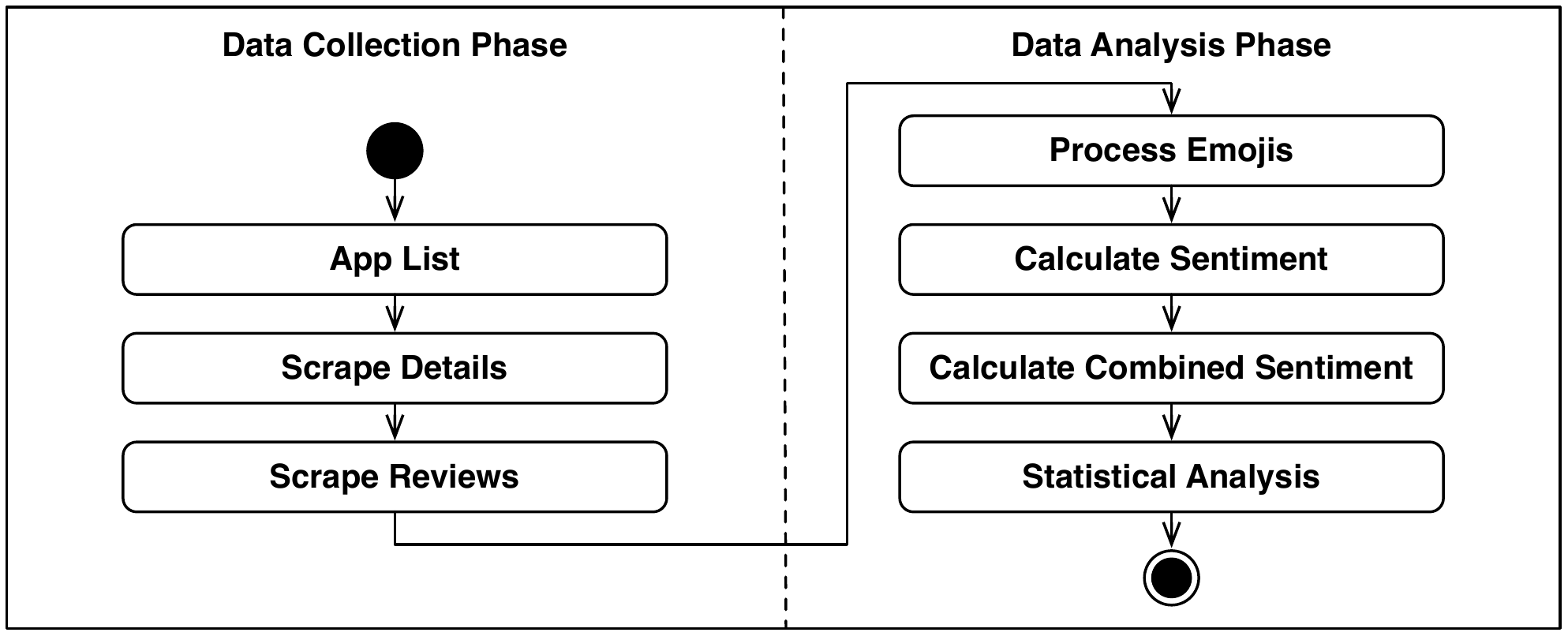}
	\caption{Research Method.}	
	\label{fig:researchmethod}
\end{figure}

The goal of this exploratory study is to find out whether and when considering the emotionality of app reviews can help to gain a deeper understanding of users' needs from the software engineering perspective. This section describes our research method and data set. Our research method consists of a data collection phase and a data analysis phase, depicted in Figure \ref{fig:researchmethod}. 
To evaluate the emotion within user reviews we measure the strength of the sentiment using \textit{SentiStrength} 2.2 with its latest configuration files from September 21, 2011. SentiStrength is a lexical sentiment extraction tool with human-level accuracy that can be applied to short informal texts written in English \cite{ASI:ASI21416}. It was trained for short social web texts such as Twitter. SentiStrength has the highest average accuracy among 15 Twitter sentiment analysis tools \cite{abbasi:2014:LREC}. Since user reviews in app stores have approximately an average length of a tweet (average length in our data set is 165) and an informal style, we chose SentiStregth over comparable tools. Alternative tools might generate other results in detail, but we assume the overall results will correlate.

\subsection{Data Collection}

For our study, we selected the top five free and paid apps (by December 18, 2016) for each of the 25 categories of the Apple AppStore. We chose the storefront of the United States to obtain reviews in English. For each app we gathered its details using the iTunes Search API\footnote{\url{https://affiliate.itunes.apple.com/resources/documentation/itunes-store-web-service-search-api}}. An app detail consists of 44 values, such as the name, version, description, category, or price. The app reviews were programmatically scraped using a self developed tool, which accesses an internal iTunes API. A review consists of 10 values, including the reviewer name, title, description, rating, or date. The update dates of apps were manually extracted from the AppStore for selected apps, as they are not available over the iTunes API. These indicate when a specific app version has been released. In addition, we collected the release notes for each version describing the introduced changes. For further analysis the data was persisted inside a MongoDB database. To enable replication, our data set is publicly available on the project website\footnote{\url{https://mast.informatik.uni-hamburg.de/app-review-analysis}}.


\subsection{Data Analysis}

\subsubsection{Process Emojis}
Emojis, a new form of emoticons, are increasingly being used by reviewers. These are unicode graphic symbols, of which 1,851 different characters exist. In contrast to emoticons their emotional content, due to their huge variety, in many cases remains unclear. To our knowledge, current sentiment analysis tools do not consider emojis. Therefore we extended SentiStrength to support emojis. Novak et al. let 83 human annotators label over 1.6 million tweets in 13 European languages by the semtiment polarity (negative, neutral, or positive) \cite{novak:Emoji:2015}. While there exists no significant difference between the languages, there is a clear difference between tweets without emojis and tweets containing emojis. The authors provide a emoji sentiment lexicon including 751 emojis giving each of them a score of either -1, 0, or 1. From the list we selected all emojis with an occurrence above 100 in the labeled tweets, resulting in 214 emojis. Using the list we replaced each of the 214 emojis with an emoticon (text representation) in order for SentiStregth to be able to consider the emojis while calculating the sentiment.

\subsubsection{Calculate Sentiment}
We provide each review (combination of title and description) as input to SentiStrength. Regarding mixed emotions \cite{10.3389/fpsyg.2015.00428}, SentiStrength returns the sentiment as two scores: a negative score from -1 (not negative) to -5 (extremely negative) and a positive score from 1 (not positive) to 5 (extremely positive). The review "I hate that u need wifi but it is great.", e.g., has a positive strength of 3 and a negative strength of -4. The sentiment is calculated by matching each token within an input to a set of dictionaries, included in the configuration files. Each dictionary defines fixed scores for specific tokens: "I hate[-4] that u need wifi but it is great[+3].". The overall score of an input is determined by using the maximum and minimum score of all annotated tokens, in this case [+3, -4]. The dictionaries also include scores for emoticons, booster words which can increase or decrease a token score ("extremely good[3] [+2 booster word]"), or lookup tables for slang words.

\subsubsection{Calculate Combined Sentiment}
Finally, we computed a combined score for each review using an approach derived from Thelwall et al. \cite{Thelwall:2012:SSD:2336247.2336261}. Considering both the negative score (n) and the positive score (p) we chose p as the combined sentiment if p + n > 0. If p + n < 0 we set n as the combined sentiment. In case p = -n and p < 4 we set the combined sentiment to 0 and assume the review to be neutral. If p = -n and p >= 4 we set the combined sentiment to undefined and the reviews are removed from the data set. In contrary to \cite{Thelwall:2012:SSD:2336247.2336261} we use a scale from -5 to 5 instead of a scale from -1 to 1. Our example review with scores [+3, -4] received a combined score of -4.


\section{Results}
\label{sec:results}


\subsection{Research Data}
\label{subsec:researchdata}

An overview of our data set is depicted in Table \ref{tab:researchdata}. We collected 7,396,551 reviews from 245 applications (125 free, 120 paid) in total. The data set consists of 23 distinct categories. The categories "Kids" and "Magazines \& Newspapers" are not present in our data set, as their top five free/paid apps were of another primary category, such as "News", "Entertainment", or "Education". Another five paid apps appeared in the top five apps of two categories, resulting in 120 paid apps in total. For each app we collected all reviews beginning with the apps' first release. The oldest feedback was provided in July, 2008. Overall, our data set spans more than 8 years.

In our data set most reviews for free apps were submitted in the category "Social Networks" (1,763,399 - 25.90\%), least in the category "Catalogs" (5,586 - 0.08\%). Most paid apps reviews were written in the category "Games" (320296 - 54.43\%), least in the category "Catalogs" (328 - 0.06\%). We could calculate the sentiment for 7,371,701 apps, only 24.850 reviews could not be classified (0.003\%)

The average length of a review (combined of title and description) is 165 characters. The average rating is 3.819 stars. The average sentiment scores are [+2.535, -1.562]. The combined sentiment average is 1.544. Both sentiments clearly highlight a more positive than negative emotion within app reviews. Only 214 reviews (0.003\%) received a vote of another app user, indicating that the provided review is helpful. 

\begin{table*}
\scriptsize
\renewcommand{\arraystretch}{1.1}
\centering
\caption{User Reviews by Categories. N = 7,396,551. {\small\textcolor{caribbeangreen}{\ensuremath\blacksquare}} = Positive, {\small\textcolor{lightgray}{\ensuremath\blacksquare}} = Neutral, {\small\textcolor{carminepink}{\ensuremath\blacksquare}} = Negative, {\small\textcolor{darkgray}{\ensuremath\blacksquare}} = Undefined.}
\label{tab:researchdata}
\begin{tabularx}{\textwidth}{lXrrrrrrl}
\toprule
\textbf{\#} & \textbf{App Category} & \textbf{\# Reviews Free Apps} & \textbf{\# Reviews Paid Apps} & \textbf{\# Reviews Sentiment} & \textbf{Mean} & \textbf{SD} & \textbf{Median} & \textbf{Distribution} \\ \midrule \rowcolor{Gray}
1  & Books 										& 111,111 	& 6,686 	& 116,947 & 1.628 & 2.437 & 3.0 & \color{caribbeangreen}\rule{2.091cm}{5pt}\color{lightgray}\rule{0.432cm}{5pt}\color{carminepink}\rule{0.455cm}{5pt}\color{darkgray}\rule{0.022cm}{5pt} \\ [0pt]
2  & Business 								& 15,539 		& 23,514 	& 38,984 & 1.878 & 2.125 & 3.0 & \color{caribbeangreen}\rule{2.245cm}{5pt}\color{lightgray}\rule{0.424cm}{5pt}\color{carminepink}\rule{0.325cm}{5pt}\color{darkgray}\rule{0.005cm}{5pt} \\ \rowcolor{Gray}
3  & Catalogs 								& 5,586 		& 328 		& 5,908 & 2.137 & 1.726 & 3.0 & \color{caribbeangreen}\rule{2.431cm}{5pt}\color{lightgray}\rule{0.394cm}{5pt}\color{carminepink}\rule{0.172cm}{5pt}\color{darkgray}\rule{0.003cm}{5pt} \\
4  & Education 								& 86,858 		& 9,046 	&	95,684 & 2.449 & 1.681 & 3.0 & \color{caribbeangreen}\rule{2.535cm}{5pt}\color{lightgray}\rule{0.318cm}{5pt}\color{carminepink}\rule{0.140cm}{5pt}\color{darkgray}\rule{0.007cm}{5pt} \\ \rowcolor{Gray}
5  & Entertainment 						& 116,613 	& 16,812 	& 132,845 & 0.619 & 2.663 & 2.0 & \color{caribbeangreen}\rule{1.527cm}{5pt}\color{lightgray}\rule{0.675cm}{5pt}\color{carminepink}\rule{0.786cm}{5pt}\color{darkgray}\rule{0.013cm}{5pt} \\
6  & Finance 									& 190,055 	& 9,303 	& 199,054 & 1.796 & 2.126 & 3.0 & \color{caribbeangreen}\rule{2.179cm}{5pt}\color{lightgray}\rule{0.474cm}{5pt}\color{carminepink}\rule{0.343cm}{5pt}\color{darkgray}\rule{0.005cm}{5pt} \\ \rowcolor{Gray}
7  & Food \& Drink 						& 297,732 	& 3,784 	& 300,848 & 2.203 & 1.893 & 3.0 & \color{caribbeangreen}\rule{2.415cm}{5pt}\color{lightgray}\rule{0.362cm}{5pt}\color{carminepink}\rule{0.216cm}{5pt}\color{darkgray}\rule{0.007cm}{5pt} \\
8  & Games 										& 189,102 	& 320,296 & 506,947 & 1.515 & 2.288 & 2.0 & \color{caribbeangreen}\rule{1.984cm}{5pt}\color{lightgray}\rule{0.600cm}{5pt}\color{carminepink}\rule{0.402cm}{5pt}\color{darkgray}\rule{0.014cm}{5pt} \\ \rowcolor{Gray}
9  & Health \& Fitness 				& 357,377 	& 16,003 	& 372,441 & 2.359 & 1.769 & 3.0 & \color{caribbeangreen}\rule{2.458cm}{5pt}\color{lightgray}\rule{0.365cm}{5pt}\color{carminepink}\rule{0.170cm}{5pt}\color{darkgray}\rule{0.008cm}{5pt} \\
10 & Lifestyle 								& 196,733 	& 4,807 	& 201,029 & 1.683 & 2.252 & 3.0 & \color{caribbeangreen}\rule{2.107cm}{5pt}\color{lightgray}\rule{0.495cm}{5pt}\color{carminepink}\rule{0.390cm}{5pt}\color{darkgray}\rule{0.008cm}{5pt} \\ \rowcolor{Gray}
11 & Medical 									& 17,054 		& 4,530 	& 21,508 & 1.932 & 2.214 & 3.0 & \color{caribbeangreen}\rule{2.285cm}{5pt}\color{lightgray}\rule{0.362cm}{5pt}\color{carminepink}\rule{0.343cm}{5pt}\color{darkgray}\rule{0.011cm}{5pt} \\
12 & Music 										& 792,204 	& 26,209 	& 815,931 & 2.150 & 2.041 & 3.0 & \color{caribbeangreen}\rule{2.383cm}{5pt}\color{lightgray}\rule{0.354cm}{5pt}\color{carminepink}\rule{0.254cm}{5pt}\color{darkgray}\rule{0.009cm}{5pt} \\ \rowcolor{Gray}
13 & Navigation 							& 293,854 	& 8,562 	& 301,625 & 2.044 & 1.947 & 3.0 & \color{caribbeangreen}\rule{2.337cm}{5pt}\color{lightgray}\rule{0.416cm}{5pt}\color{carminepink}\rule{0.240cm}{5pt}\color{darkgray}\rule{0.008cm}{5pt} \\
14 & News 										& 352,659 	& 20,565 	& 372,258 & 1.121 & 2.488 & 2.0 & \color{caribbeangreen}\rule{1.824cm}{5pt}\color{lightgray}\rule{0.588cm}{5pt}\color{carminepink}\rule{0.580cm}{5pt}\color{darkgray}\rule{0.008cm}{5pt} \\ \rowcolor{Gray}
15 & Photo \& Video 					& 601,687 	& 45,637 	& 642,951 & 0.429 & 2.829 & 0.0 & \color{caribbeangreen}\rule{1.480cm}{5pt}\color{lightgray}\rule{0.624cm}{5pt}\color{carminepink}\rule{0.876cm}{5pt}\color{darkgray}\rule{0.020cm}{5pt} \\
16 & Productivity 						& 127,440 	& 9,796 	& 136,828 & 1.382 & 2.368 & 2.0 & \color{caribbeangreen}\rule{1.952cm}{5pt}\color{lightgray}\rule{0.565cm}{5pt}\color{carminepink}\rule{0.474cm}{5pt}\color{darkgray}\rule{0.009cm}{5pt} \\ \rowcolor{Gray}
17 & Reference 								& 505,966 	& 20,145 	& 525,456 & 2.736 & 1.334 & 3.0 & \color{caribbeangreen}\rule{2.728cm}{5pt}\color{lightgray}\rule{0.197cm}{5pt}\color{carminepink}\rule{0.071cm}{5pt}\color{darkgray}\rule{0.004cm}{5pt} \\
18 & Shopping 								& 129,449 	& 2,192 	& 131,227 & 2.022 & 2.181 & 3.0 & \color{caribbeangreen}\rule{2.301cm}{5pt}\color{lightgray}\rule{0.366cm}{5pt}\color{carminepink}\rule{0.324cm}{5pt}\color{darkgray}\rule{0.009cm}{5pt} \\ \rowcolor{Gray}
19 & Soc. Networking 				& 1,763,399 & 1,970 	& 1,759,810 & 0.993 & 2.634 & 2.0 & \color{caribbeangreen}\rule{1.761cm}{5pt}\color{lightgray}\rule{0.572cm}{5pt}\color{carminepink}\rule{0.657cm}{5pt}\color{darkgray}\rule{0.009cm}{5pt} \\
20 & Sports 									& 174,046 	& 3,483 	& 177,055 & 0.959 & 2.617 & 2.0 & \color{caribbeangreen}\rule{1.777cm}{5pt}\color{lightgray}\rule{0.547cm}{5pt}\color{carminepink}\rule{0.668cm}{5pt}\color{darkgray}\rule{0.008cm}{5pt} \\ \rowcolor{Gray}
21 & Travel 									& 139,285 	& 3,373 	& 142,278 & 1.442 & 2.360 & 3.0 & \color{caribbeangreen}\rule{1.958cm}{5pt}\color{lightgray}\rule{0.541cm}{5pt}\color{carminepink}\rule{0.493cm}{5pt}\color{darkgray}\rule{0.008cm}{5pt} \\
22 & Utilities 								& 106,584 	& 9,522 	& 115,710 & 1.305 & 2.426 & 2.0 & \color{caribbeangreen}\rule{1.905cm}{5pt}\color{lightgray}\rule{0.573cm}{5pt}\color{carminepink}\rule{0.511cm}{5pt}\color{darkgray}\rule{0.010cm}{5pt} \\ \rowcolor{Gray}
23 & Weather 									& 237,739 	& 21,916 	& 258,377 & 1.273 & 2.463 & 2.0 & \color{caribbeangreen}\rule{1.933cm}{5pt}\color{lightgray}\rule{0.523cm}{5pt}\color{carminepink}\rule{0.529cm}{5pt}\color{darkgray}\rule{0.015cm}{5pt} \\ \midrule

   &													& $\sum 6,808,072$ & $\sum 588,479$ & $\sum 7,371,701$ & $\varnothing 1.845$ & $\varnothing 2.211$ & $\varnothing 2.522$ & \\

\bottomrule
\end{tabularx}
\end{table*}


\subsection{Emotion and the Rating}
\label{sec:rating}
\begin{figure}[hbt!]
	\includegraphics[width=0.5\textwidth]{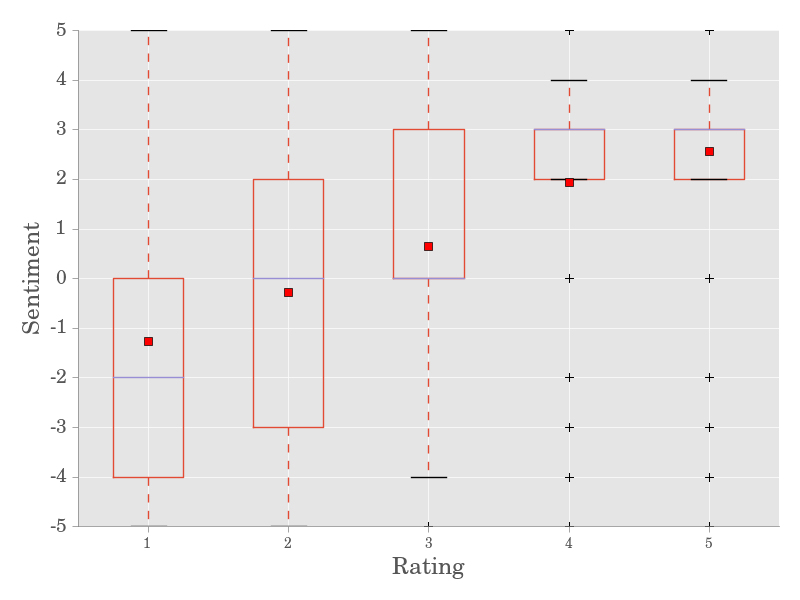}
	\caption{Sentiment per Star Rating.}	
	\label{fig:sentstars}
\end{figure}

Besides the review, a user can give a star rating on a scale from 1 to 5 stars. We assume that a higher star rating correlates with stronger positive emotions and vice versa, that a low star rating correlates with a negative emotion. Figure \ref{fig:sentstars} shows a box plot of the sentiment for each star rating. At a first glance the plot seems to confirm our assumption. The Pearson correlation coefficient of 0.5699 also draws attention towards a weak positive linear correlation. Although this finding is not surprising, since it follows the intuitional assumption, we will take a closer look at the outliers. Even though the Spearman rank correlation is influenced less by outliers, it also draws a weak correlation (0.560776). Especially for the four and five star ratings, many reviews are classified as outliers. Out of 612,271 reviews rated with five stars, 199,194 have a combined score of -4 or -5 (32.53\%). This can be explained by the nature of SentiStrength: "I would be very sad without it" has a rating of five stars, but the sentiscore  has positive strength 1 and negative strength -5, as a consequence of the booster word "very" and the negative word "sad". 

This shows that on average sentiment and the star rating are correlated, but are not a sound measure to determine if a user is happy with the app. This affects studies that rely on the link between emotion and user satisfaction. 


\subsection{Emotion and the Price}
\label{sec:money}
We raise the question if users react emotionally stronger when money is involved. Does a user react with more emotion to a bug or a new added feature when the app was subject to a charge, and does the price make a difference? Harman et al. \cite{harman:2012:MSR} showed that there is no correlation between price and downloads, nor between price and rating. Pagano and Maalej \cite{pagano:RE:2013} found that there is significant increase in feedback length between lower-price and higher-price applications, which we can confirm based on our sample. We can reject the hypothesis that there is a significant linear correlation between price and emotion (Pearson/Spearman: 0.021/0.026). A possible explanation might be that users, who are generally willing to provide feedback do not differentiate between free, low-price, and higher-priced apps and give feedback based on an intrinsic motivation. Their inclination of expression stays the same, independent of the pricing.  


\subsection{Emotion and the Content}
\label{sec:content}

\begin{figure}[hbt!]
	\includegraphics[width=0.5\textwidth]{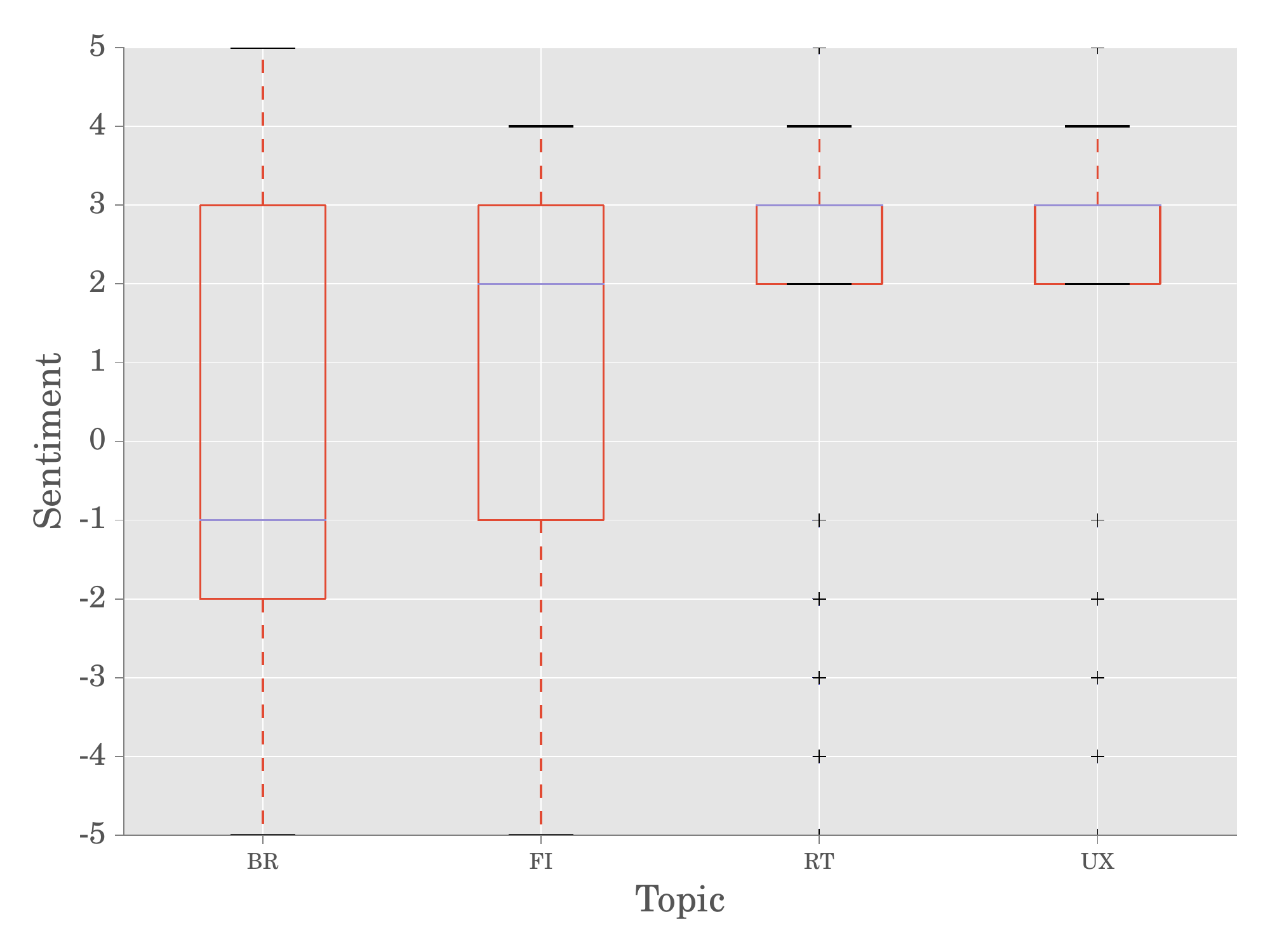}
	\caption{Sentiment per Topic.}	
	\label{fig:04-cluster}
\end{figure}

Pagano and Maalej \cite{pagano:RE:2013} described 17 different topics in user reviews. Especially for software engineers it is important to filter by the topics \textit{Bug Reports}, \textit{Feature Requests} and \textit{User Experience}. Maalej and Nabil \cite{maalej:RE:2015} compared the accuracy of different techniques for the classification into the topics: \textit{Bug Reports}, \textit{Feature Requests}, \textit{User Experience} and \textit{Rating}. A rating here is a literal repetition of the star rating including praise, dispraise, a distractive critique, or a dissuasion.
The authors used \textit{bag of words} with and without lemmatization and stopword removal, the star rating, tense and the sentiment. They trained a classifier with different combinations of these features. The authors found that there is not a one-size-fits-all solution. The sentiment could be used to achieve a higher accuracy in some cases. At this, it did not make a difference if they used the two sided sentiment or the combined sentiment. However, not in every case could the use of the sentiment as an additional feature could lead to better results. Maalej et al. \cite{maalej:2016:JRE} showed that other features are more informative for the classification. Only for the topics \textit{User Experience} and \textit{Bug Report}, the sentiment was ranked amongst the 10 most important features.
We analyzed the sentiment based on their labeled dataset and found that on average a bug report has a more negative sentiment than a user experience (cp. Figure \ref{fig:04-cluster}). A closer look on the dispersion measures reveals that the sentiments are very scattered, especially for the topics \textit{Bug Report} and \textit{Feature Request} (cp. Table \ref{Dispersion}). \textit{Ratings} and \textit{User Experience} have a very high amount of outliers.

\begin{table}[hbt]
\scriptsize
\renewcommand{\arraystretch}{1.1}
\centering
\caption{Dispersion of Sentiments.}
\begin{tabularx}{\columnwidth}{Xrrrr}
\toprule
\textbf{Topic}				&	\textbf{\#} \textbf{Reviews} & \textbf{Range}    & \textbf{IQR (Q\textsubscript{.75} - Q\textsubscript{.25})}	& \textbf{SD}\\ \midrule \rowcolor{Gray}
Bug Report            & 	379		& 10              & 5					&  2.62        \\
Feature Request       & 	295		& 9               & 4          			&  2.35       \\ \rowcolor{Gray}
User Experience      & 	735		& 10              & 1   					&  1.99           \\
Rating             	  & 	2,604		& 10              & 1          			&  1.90           \\
 \bottomrule
\end{tabularx}
\label{Dispersion}
\end{table}
  
We think that the sentiment can be a much more important informative feature when the SentiStrenght tool is adjusted. Since SentiStrength was not intended for software engineering purposes, words that have a relative negative sentiment in the domain of software engineering are not necessarily negative by their nature. For example, the fictitious user review \textit{"This app is really buggy and crashes all the time on my phone"} would receive a neutral sentiment [-1, 1]. In the software engineering domain, words like \textit{bug} or \textit{crash} have a very negative connotation.


\subsection{Patterns in Emotions}
\label{sec:patterns}

The emotionality of app reviews can vary over time. Reasons for this can be manifold, e.g., features are added, bugs occur, or feature requests are ignored by app developers.

\begin{figure}[t]
	\includegraphics[width=0.5\textwidth]{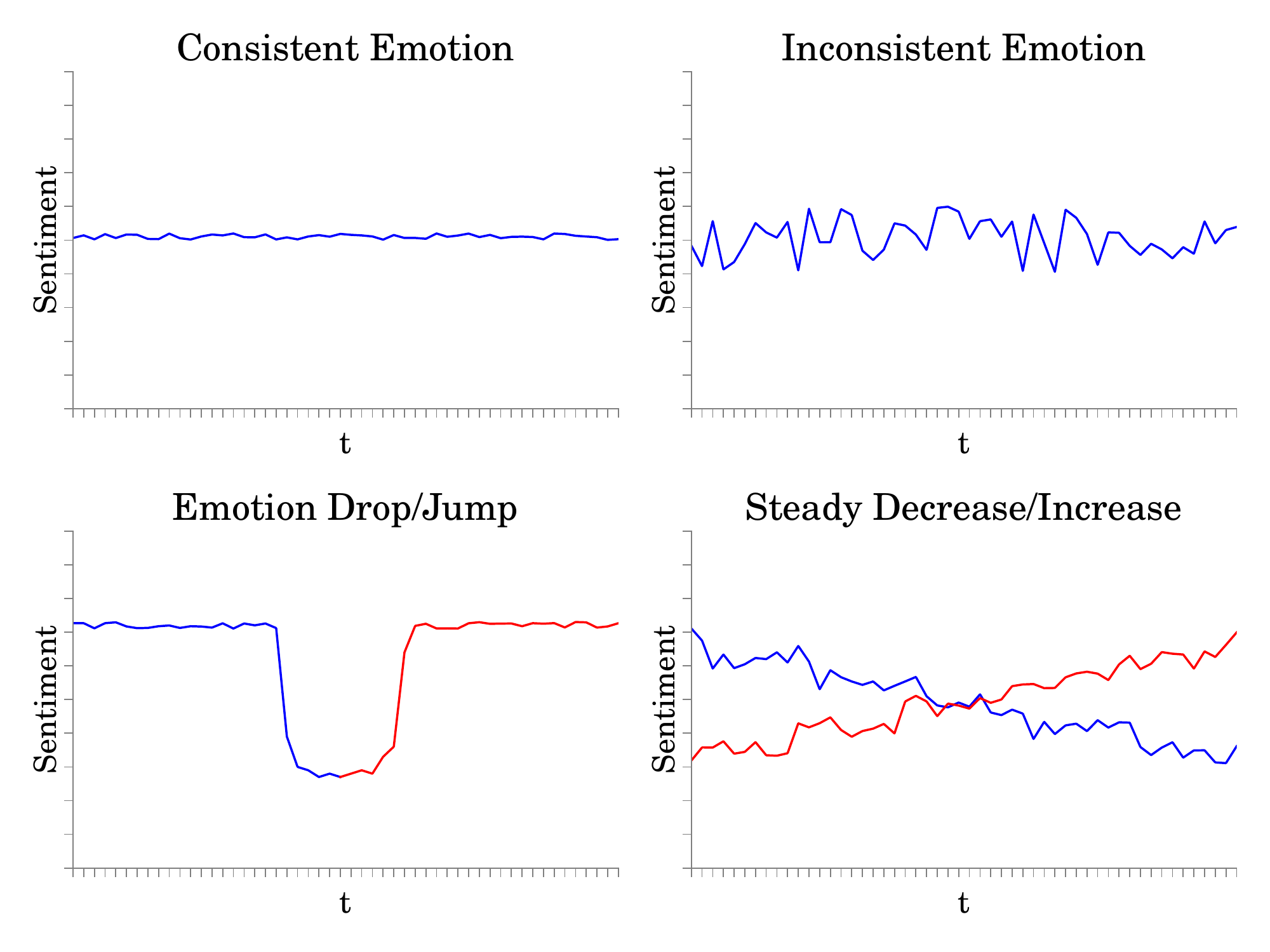}
	\caption{Reoccurring Patterns in Emotions.}	
	\label{fig:patternsinemotions}
\end{figure}

To understand the development of emotion, we selected all apps of our data set having more than 1,000 reviews within the timeframe from January 4, 2016 to December 18, 2016. For each app, we plotted the sentiment for the given timeframe. In these plots we found six reoccurring patterns, depicted in Figure \ref{fig:patternsinemotions}. First, \textit{Consistent Emotion} with a stable sentiment. Second, \textit{Inconsistent Emotion}, which is less constant but does not show a clear positive or negative trend. Third, \textit{Sentiment Drops} that show a sudden decrease. Fourth, \textit{Sentiment Jumps} where the sentiment shows a sudden positive increase. The fifth and sixth pattern, \textit{Steady Decrease/Increase}, show a constant negative or positive trend. Whereas the first two patterns do not have additional value for the developers, the observation of the latter can give additional insights. Particularly, when emotional drops or jumps occur after app updates, developers quickly receive feedback regarding the introduced changes.

We manually classified 101 plots of the evolution of emotion over one year (grouped by weekly average) into the aforementioned classes. When multiple patterns during the time period occured, we classified them into multiple classes. Most of the plots (62\%) showed a inconsistent emotional pattern. 15\% had a significant jump in emotion and 9\% showed a drop, while 10\% had a steady decrease, and only 3\% a steady increase. 15\% contained a consistent high or low emotion. 

We took a detailed look into the reviews of two apps within the categories \textit{emotion jump} and \textit{emotion drop} to better understand the emergence of such patterns. We assume these categories to be the most interesting for developers. Figure \ref{fig:bankofamerica} shows the evolution of the sentiment for the "Bank of America" app, while Figure \ref{fig:gmail} shows the sentiment for the "Gmail" app. In both figures we highlighted the release of app updates using grey vertical lines.


\subsubsection{Bank of America} 
\begin{figure}[t]
	\includegraphics[width=0.5\textwidth]{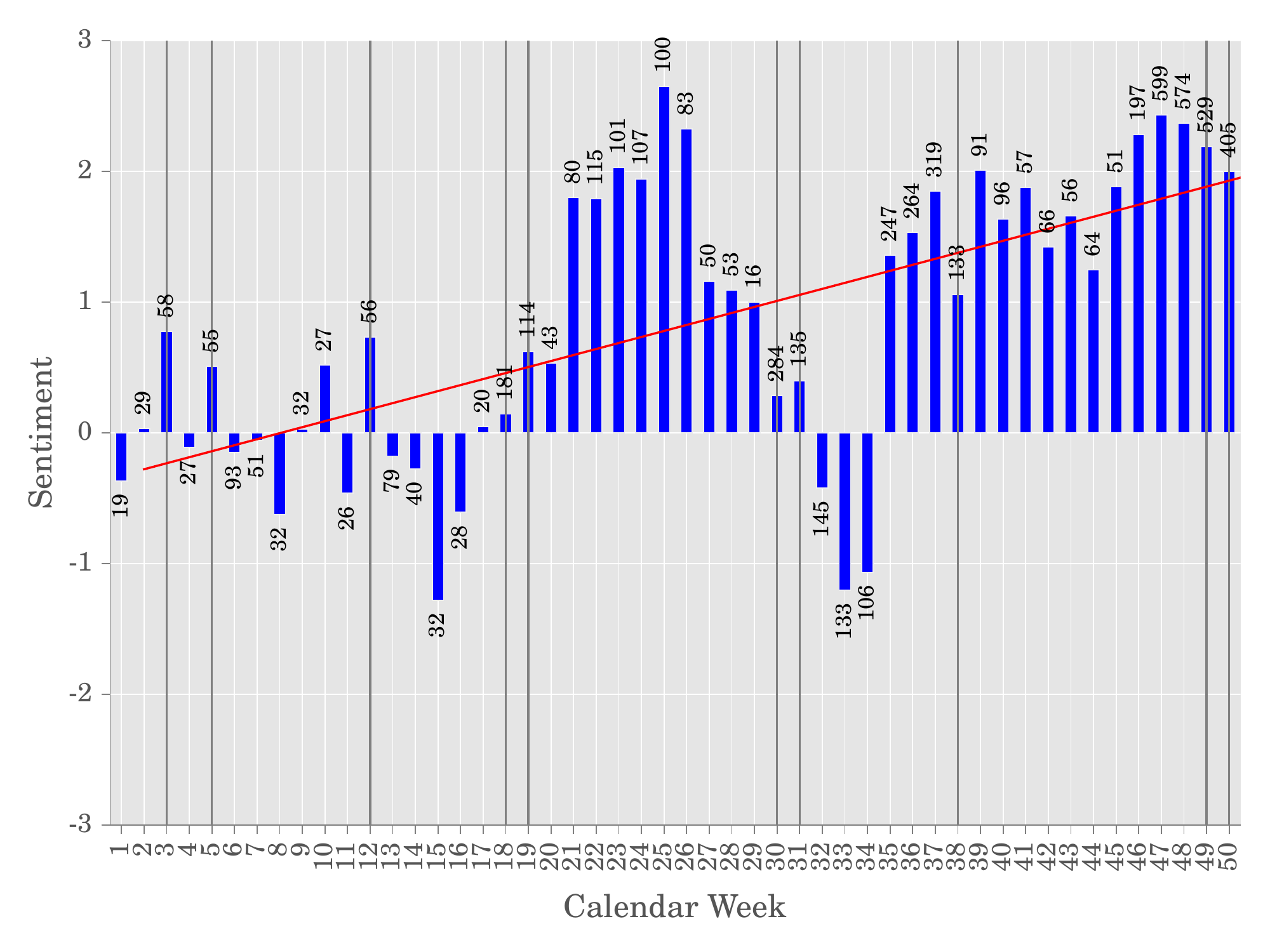}
	\caption{Sentiment in 2016 of "Bank of America" App.}	
	\label{fig:bankofamerica}
\end{figure}

For this app 6,307 reviews were submitted in 2016. Figure \ref{fig:bankofamerica} can be separated into five sections. Section 1 ranges from week 1 to week 12 with an inconsistent sentiment oscillating around 0. App updates including minor improvements released in week 3, 5 and 12 temporarily raise the sentiment above 0, afterwards the sentiment becomes negative again. The reviewers have extremely contrary positions, while some write \textit{"Good App!"} others report \textit{"I rather run 10 miles to a branch, barefoot on snow, than use this app."}. Section 2 (week 13 - 16) is initiated by a new app update and has a sentiment below 0. Mostly iPhone 6s users report that the app crashes. In the beginning of section 3 (week 17 - 31) two app updates are released, which introduce the support for iOS 9 (preinstalled on iPhone 6s). As a result, the sentiment increases with the most positive reviews in week 25, overall the sentiment in this section is positive in every week. Section 4 (week 32 - 34) has a negative sentiment. It is initiated by two app updates that introduce a redesign of the app and a minimum required iOS version of 8. Users mostly state \textit{"iOS 8 now required! [...]"}, as a consequence they report \textit{"Doesn't work with iPhone 4 [...]"}. In the last section (week 35 - 50) the sentiment is above 0 again. In week 38 an app update is released, which introduces new features and causes the sentiment to further increase. Overall the reviews become shorter and less informative, mostly depicting the subjective impression of the users such as \textit{"Solid app"}, \textit{"Love it!!!"}, or \textit{"Great app"}.


\subsubsection{Gmail} 
\begin{figure}[t]
	\includegraphics[width=0.5\textwidth]{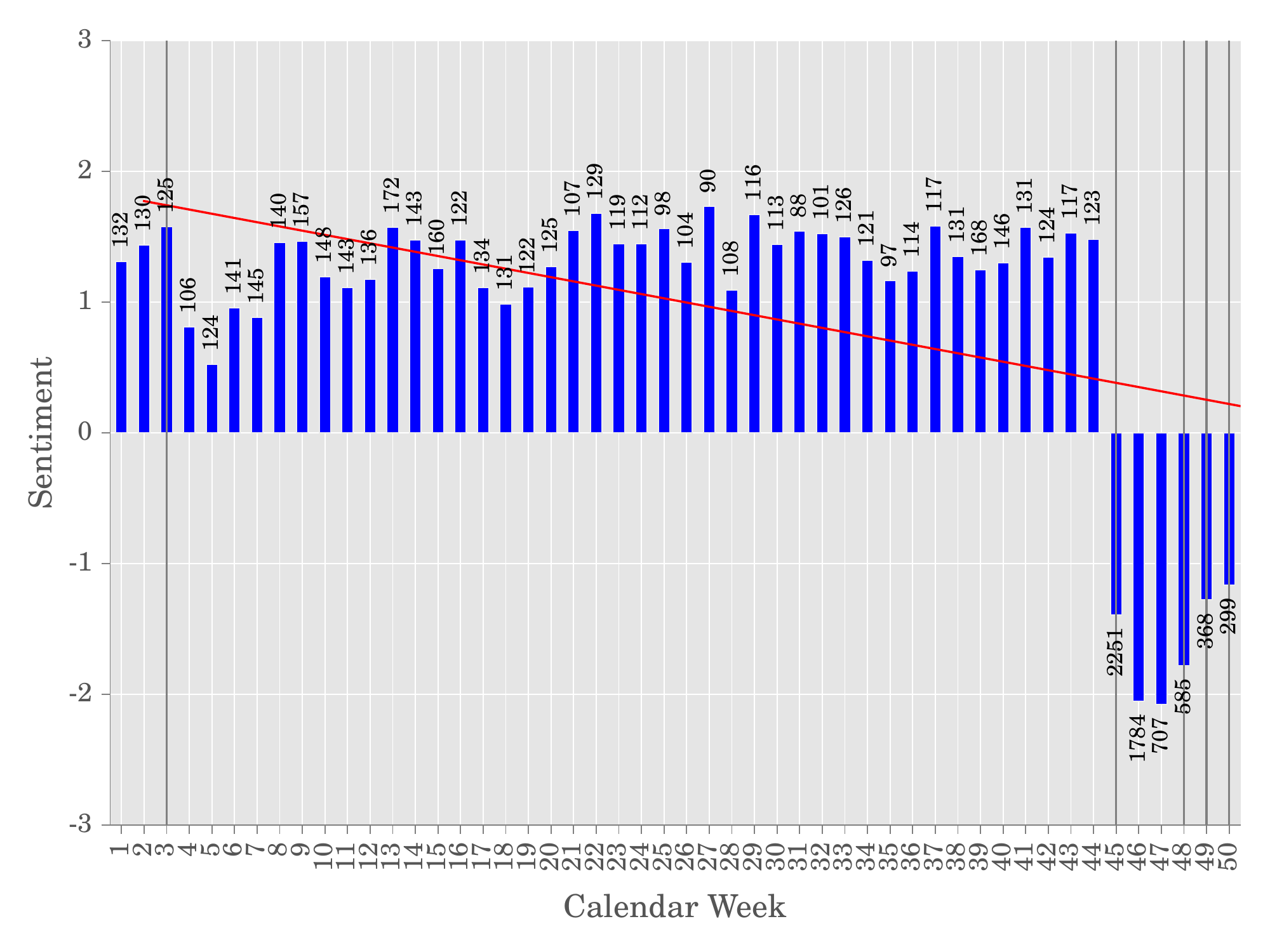}
	\caption{Sentiment in 2016 of "Gmail" App.}	
	\label{fig:gmail}
\end{figure}

In 2016 users submitted 11,629 reviews for the app. Figure \ref{fig:gmail} can be divided into two sections. The first section ranges from week 1 to week 44 with an ongoing positive sentiment. The second section ranges from week 45 to week 50 with an enduring negative sentiment. In section 1 users mostly submit positive reviews, such as \textit{"This app does it all!"}, with an average length of 145 characters. In this section there is only one remarkably low sentiment in week 5 where users report issues when loading and deleting emails. We expect these issues to be caused by server failures, as there were no app updates released. Section 2 begins with the release of a redesigned app update. Overall, users submit 6,104 reviews in this section (1,017 reviews/week on average) compared to section 1 with 5,525 reviews (126 reviews/week on average). The average length of the reviews also increases to 273 characters, nearly twice as long as in section 1. Until week 47 the sentiment becomes more negative, users mostly complain about features that have been changed \textit{"When Deleting messages now requires an extra step - you failed!"}, or features that have been removed such as \textit{"Bring back Mark as Unread. Why on earth would you remove this feature?"} and \textit{"The new app no longer allows me to select multiple emails."}. In week 48 the app developers react with weekly updates integrating features as requested by the users, e.g., version 5.0.6 (November 30, 2016): \textit{"Select multiple messages [...]"}, and \textit{"Mark as read/unread: [...]"}. Starting with this release the sentiment becomes less negative.

Summarizing our results, it can be said that app updates have the most influence on the emotion in app reviews. Integrating single features as requested by users positively influences the sentiment, while, e.g., releasing a completely redesigned app can cause the previously positive sentiment to turn into a negative sentiment. This also applies for changing apps' requirements such as the minimum supported iOS version. When providing negative reviews users write longer and more detailed reviews. After releasing major updates the number of reviews per week also increases for a specific period of time.

\section{Discussion and Future Directions}
\label{sec:conclusion}

We found that emotion in user reviews can be seen as a meaningful additional meta-data. The fact that we could only observe weak correlations with other indicators tempers the expectations of current approaches. This can be linked back to the general approach of tools like SentiStrength. In future work  it will be necessary to adjust tools and approaches to better fit the contents in user reviews. The correlation between the emotional sentiment and the user ratings is only weakly defined. Especially for very positive ratings with four and five stars, the rating itself is a better indicator for the users' satisfaction than the sentiment. The emotional sentiment is not influenced by the price. While reviews for higher priced apps tend to be longer on average, users do not react with more emotion, neither in a positive nor in a negative way. We assume that the price paid for an app does not give additional incentives to review apps and that users who are willing to provide feedback have other motivations, which are not based on the price. The sentiment becomes more relevant when user reviews are classified based on their topic. We see a potential to further increase the quality of classification techniques when tools, such as SentiStrength, are adjusted to software topics by applying different sentiment weights to given buzzwords such as "bug" or "crash". A promising insight is the development of the sentiment over time and the reoccurring patterns that could be found in our analysis. Particularly mixed emotions attracted our interest, since the manual analysis indicates that users actively discuss pros and cons of a release. We assume that adding additional features to the timeline, we can give more exact assumptions about the users perception of a release. Our initial findings motivate to adjust sentiment analysis towards a better fit for the purpose of software reviews. In this study we used SentiStrength knowingly without further modification. We assume that especially the emotional weighting of terms must be adjusted. 

In this initial study we found that adding the emotion of a review as a feature can provide additional information. Whereas some of our findings confirmed intuitive assumptions, others revealed that the calculated sentiment often diverges.


\bibliographystyle{abbrv}
\bibliography{SEmotion_2017}

\begin{thebibliography}{10}

\bibitem{abbasi:2014:LREC}
A.~Abbasi, A.~Hassan, and M.~Dhar.
\newblock Benchmarking twitter sentiment analysis tools.
\newblock In {\em Ninth Language Resources and Evaluation Conference.}, pages
  823--829, 2014.

\bibitem{10.3389/fpsyg.2015.00428}
R.~Berrios, P.~Totterdell, and S.~Kellett.
\newblock Eliciting mixed emotions: a meta-analysis comparing models, types,
  and measures.
\newblock {\em Frontiers in Psychology}, 6:428, 2015.

\bibitem{Chen:2014:AMI:2568225.2568263}
N.~Chen, J.~Lin, S.~C.~H. Hoi, X.~Xiao, and B.~Zhang.
\newblock Ar-miner: Mining informative reviews for developers from mobile app
  marketplace.
\newblock In {\em Proceedings of the 36th International Conference on Software
  Engineering}, ICSE 2014, pages 767--778, New York, NY, USA, 2014. ACM.

\bibitem{Fu:ICKDDM:2013}
B.~Fu, J.~Lin, L.~Li, C.~Faloutsos, J.~Hong, and N.~Sadeh.
\newblock Why people hate your app: Making sense of user feedback in a mobile
  app store.
\newblock In {\em Proceedings of the 19th ACM SIGKDD International Conference
  on Knowledge Discovery and Data Mining}, KDD '13, pages 1276--1284, New York,
  NY, USA, 2013. ACM.

\bibitem{Goul:2012:HICSS}
M.~Goul, O.~Marjanovic, S.~Baxley, and K.~Vizecky.
\newblock Managing the enterprise business intelligence app store: Sentiment
  analysis supported requirements engineering.
\newblock In {\em 2012 45th Hawaii International Conference on System
  Sciences}, pages 4168--4177, Jan 2012.

\bibitem{Guzman:2014:SAC}
E.~Guzman, D.~Az\'{o}car, and Y.~Li.
\newblock Sentiment analysis of commit comments in github: An empirical study.
\newblock In {\em Proceedings of the 11th Working Conference on Mining Software
  Repositories}, MSR 2014, pages 352--355, New York, NY, USA, 2014. ACM.

\bibitem{Guzman:RE:2014}
E.~Guzman and W.~Maalej.
\newblock How do users like this feature? a fine grained sentiment analysis of
  app reviews.
\newblock In {\em Proceedings of the 22nd RE Conference}, 2014.

\bibitem{Harman:MSR:2012}
M.~Harman, Y.~Jia, and Y.~Zhang.
\newblock App store mining and analysis: Msr for app stores.
\newblock In {\em Mining Software Repositories (MSR), 2012 9th IEEE Working
  Conference on}, pages 108--111, June 2012.

\bibitem{harman:2012:MSR}
M.~Harman, Y.~Jia, and Y.~Zhang.
\newblock App store mining and analysis: Msr for app stores.
\newblock In {\em 2012 9th IEEE Working Conference on Mining Software
  Repositories (MSR)}, pages 108--111, June 2012.

\bibitem{Iacob:MSR:2013}
C.~Iacob and R.~Harrison.
\newblock Retrieving and analyzing mobile apps feature requests from online
  reviews.
\newblock In {\em Proceedings of the 10th Working Conference on Mining Software
  Repositories}, MSR '13, pages 41--44, Piscataway, NJ, USA, 2013. IEEE Press.

\bibitem{Jongeling:2015:ICSME}
R.~Jongeling, S.~Datta, and A.~Serebrenik.
\newblock Choosing your weapons: On sentiment analysis tools for software
  engineering research.
\newblock In {\em 2015 IEEE International Conference on Software Maintenance
  and Evolution (ICSME)}, pages 531--535, Sept 2015.

\bibitem{Jongeling:2017:ESE}
R.~Jongeling, P.~Sarkar, S.~Datta, and A.~Serebrenik.
\newblock On negative results when using sentiment analysis tools for software
  engineering research.
\newblock {\em Empirical Software Engineering}, pages 1--42, 2017.

\bibitem{li:2010:user}
H.~Li, L.~Zhang, L.~Zhang, and J.~Shen.
\newblock A user satisfaction analysis approach for software evolution.
\newblock In {\em Progress in Informatics and Computing (PIC), 2010 IEEE
  International Conference on}, volume~2, pages 1093--1097. IEEE, 2010.

\bibitem{maalej:2016:JRE}
W.~Maalej, Z.~Kurtanovi\'{c}, H.~Nabil, and C.~Stanik.
\newblock On the automatic classification of app reviews.
\newblock {\em Requiruirements Engineering}, 21(3):311--331, Sept. 2016.

\bibitem{maalej:RE:2015}
W.~Maalej and H.~Nabil.
\newblock Bug report, feature request, or simply praise? on automatically
  classifying app reviews.
\newblock In {\em Requirements Engineering Conference (RE), 2015 IEEE 23rd
  International}, pages 116--125. IEEE, 2015.

\bibitem{novak:Emoji:2015}
P.~K. Novak, J.~Smailovi{\'c}, B.~Sluban, and I.~Mozeti{\v{c}}.
\newblock Sentiment of emojis.
\newblock {\em PloS one}, 10(12):e0144296, 2015.

\bibitem{Novielli:2014:SSE}
N.~Novielli, F.~Calefato, and F.~Lanubile.
\newblock Towards discovering the role of emotions in stack overflow.
\newblock In {\em Proceedings of the 6th International Workshop on Social
  Software Engineering}, SSE 2014, pages 33--36, New York, NY, USA, 2014. ACM.

\bibitem{pagano:RE:2013}
D.~Pagano and W.~Maalej.
\newblock User feedback in the appstore: An empirical study.
\newblock In {\em 21st IEEE International Requirements Engineering
  Conference,}, pages 125--134. IEEE, 2013.

\bibitem{Thelwall:2012:SSD:2336247.2336261}
M.~Thelwall, K.~Buckley, and G.~Paltoglou.
\newblock Sentiment strength detection for the social web.
\newblock {\em J. Am. Soc. Inf. Sci. Technol.}, 63(1):163--173, Jan. 2012.

\bibitem{ASI:ASI21416}
M.~Thelwall, K.~Buckley, G.~Paltoglou, D.~Cai, and A.~Kappas.
\newblock Sentiment strength detection in short informal text.
\newblock {\em Journal of the American Society for Information Science and
  Technology}, 61(12):2544--2558, 2010.

\bibitem{Turney:2002:Corr}
P.~D. Turney.
\newblock Thumbs up or thumbs down? semantic orientation applied to
  unsupervised classification of reviews.
\newblock {\em CoRR}, cs.LG/0212032, 2002.

\end{thebibliography}


\end{document}